\documentclass[conference]{IEEEtran}
\IEEEoverridecommandlockouts
\usepackage{cite}
\usepackage{amsmath,amssymb,amsfonts}
\usepackage{algorithmic}
\usepackage{graphicx}
\usepackage{textcomp}
\usepackage{multirow}
\usepackage{xcolor}
\usepackage{csquotes}
\usepackage{url}

\usepackage{etoolbox}
\AtBeginEnvironment{quote}{\par\singlespacing\small}
\def\BibTeX{{\rm B\kern-.05em{\sc i\kern-.025em b}\kern-.08em
    T\kern-.1667em\lower.7ex\hbox{E}\kern-.125emX}}

\begin{document}
\title{Analyzing Challenges in Deployment of the SLSA Framework for Software Supply Chain Security}

\author{\IEEEauthorblockN{Mahzabin Tamanna}
\IEEEauthorblockA{\textit{North Carolina State University}\\
\textit{Raleigh, NC, USA} \\
\textit{mtamann@ncsu.edu}}
\and
\IEEEauthorblockN{Sivana Hamer}
\IEEEauthorblockA{\textit{North Carolina State University} \\
\textit{Raleigh, NC, USA} \\
\textit{sahamer@ncsu.edu}}

\and
\IEEEauthorblockN{{Mindy Tran}
\IEEEauthorblockA{\textit{Max Planck Institute for Security and Privacy} \\
Bochum, Germany\\
mindy.tran@mpi-sp.org}}
\and
\IEEEauthorblockN{Sascha Fahl}
\IEEEauthorblockA{\textit{CISPA Helmholtz Center for } \\
 \textit{Information
Security, Hanover, Germany}\\
 \textit{sascha.fahl@cispa.de}}
\and
\IEEEauthorblockN{Yasemin Acar}
\IEEEauthorblockA{\textit{Paderborn University}\\
\textit{The George Washington University}\\
Paderborn, Germany and Wahington, USA \\
yasemin.acar@uni-paderborn.de}\\
\and
\IEEEauthorblockN{Laurie Williams}
\IEEEauthorblockA{\textit{North Carolina State University} \\
Raleigh, NC, USA \\
lawilli3@ncsu.edu}
}

\maketitle

\begin{abstract}
In 2023, Sonatype reported a 200\% increase in software supply chain attacks, including major build infrastructure attacks. 
To secure the software supply chain, practitioners can follow security framework guidance like the Supply-chain Levels for Software Artifacts (SLSA). However, recent surveys and industry summits have shown that despite growing interest, the adoption of SLSA is not widespread. To understand adoption challenges, \textit{the goal of this study is to aid framework authors and practitioners in improving the adoption and development of Supply-Chain Levels for Software Artifacts (SLSA) through a qualitative study of SLSA-related issues on GitHub}. 
We analyzed 1,523 SLSA-related issues extracted from 233 GitHub repositories. We conducted a topic-guided thematic analysis, leveraging the Latent Dirichlet Allocation (LDA) unsupervised machine learning algorithm, to explore the challenges of adopting SLSA and the strategies for overcoming these challenges. We identified four significant challenges and five suggested adoption strategies. The two main challenges reported are complex implementation and unclear communication, highlighting the difficulties in implementing and understanding the SLSA process across diverse ecosystems. The suggested strategies include streamlining provenance generation processes, improving the SLSA verification process, and providing specific and detailed documentation. Our findings indicate that some strategies can help mitigate multiple challenges, and some challenges need future research and tool enhancement.
\end{abstract}
\vspace{10pt}
\begin{IEEEkeywords}
SLSA, Software supply-chain security, Security Framework
\end{IEEEkeywords}

\section{Introduction}
Software supply chain attacks are predicted to cost businesses \$138 billion globally by 2031~\cite{Cybercrimemagazine}. In 2023, Sonatype reported that 245,032 malicious open-source packages were downloaded, reflecting a 200\% increase in software supply chain attacks~\cite{Sonatype-23}. Well-known build infrastructure attacks, like SolarWinds and Codecov, affected thousands of customers and hundreds of businesses and government agencies~\cite{Comparitech}. Consequently, the US Executive Order (EO) 14028 ~\cite{EO} highlights the urgent need to improve transparency and integrity in open-source artifacts to enhance supply chain security.
The Open Source Security Foundation (OpenSSF) introduced a security framework, Supply-chain Levels for Software Artifacts (SLSA)~\cite{slsa} (described in Sec.~\ref{SLSA}). SLSA is designed to aid organizations in securing software and infrastructure development and deployment by preventing tampering and ensuring integrity. Organizations can leverage SLSA to implement artifact management practices and automate decisions to protect the software supply chain~\cite{legitsecurity,bytesafe,slsa-googleblog}. Software supply chain attacks are increasing. For example, in March 2024, xz-utils, a data compression library in Linux distributions, was compromised with malicious commits and introduced a backdoor affecting SSH as a dependency~\cite{freund2024}. Implementing SLSA practices, such as source integrity verification, software provenance tracking, secure builds, and dependency verification, could help reduce the risk of software supply chain attacks.
Several studies have highlighted the advantages of SLSA in securing software supply chain~\cite{melara2022software,tran2023toward,ladisa2023journey}, but adopting SLSA poses challenges. Based on an industry summit and insights from 19 practitioners from 17 various companies, Tran et al. ~\cite{tran2023s3c2} reported that practitioners consider SLSA promising for securing systems but face limitations in integrating it into their ecosystems. Practitioners also have uncertainty about overcoming obstacles in securing buildings using SLSA. According to an analysis of the 90 projects from Java, Python, and npm ecosystems, most of the top projects across the ecosystems struggle to fulfill the core requirement of SLSA~\cite{hassanshahi2023macaron}. A joint survey conducted by the OpenSSF, Chainguard, Eclipse, and the Rust foundation in 2023 ~\cite{slsa-openssf} revealed that over 50\% of participants find it difficult to implement certain SLSA practices, such as hermetic builds. Moreover, survey respondents stated that making the provenance document available is a complex practice for organizations. Therefore, systematically analyzing the SLSA-related challenges and proposed strategies to overcome the challenges can help in increasing the adoption of security frameworks like SLSA.

    \textit{The goal of this study to aid framework authors and practitioners in improving the adoption and development of Supply-Chain Levels for Software Artifacts (SLSA) through a qualitative study of SLSA-related issues on GitHub.}

In this study, we address the following research questions:
\begin{itemize}

    \item \textbf{RQ1:} What challenges do practitioners encounter while deploying SLSA?
    \item \textbf{RQ2:} What strategies do software practitioners suggest to framework authors for increasing SLSA adoption?
        
\end{itemize}

In this study, \textit{framework authors} refers to those who developed or maintain the SLSA framework; \textit{practitioners} refers to people who created issues for SLSA-related challenges, provided strategies to overcome them, or involved in the discussion on GitHub.

To address our research questions, we analyzed 1,523 SLSA-related issues from 233 GitHub repositories across June 2021 to September 2023. We used unsupervised machine learning, Latent Dirichlet Allocation (LDA), and qualitative analysis. Framework authors can improve framework usability and adoption based on our findings. Practitioners can gain an understanding of challenges and solutions to better secure their software supply chains. In addition, addressing these adoption challenges may enhance overall software ecosystem security, benefiting software development and deployment stakeholders. The contributions of our study are below:

 \begin{enumerate}
    \item A set of challenges practitioners encountered in adopting SLSA. 
    \item A set of strategies discussed by practitioners to overcome the challenges of deploying SLSA.
    \item A set of recommendations based on our work for security framework authors, practitioners, and researchers to improve the adoption of security frameworks like SLSA.
   
\end{enumerate}
\section{SUPPLY-CHAIN LEVELS FOR SOFTWARE
ARTIFACTS (SLSA)}\label{SLSA}
Supply Chain Levels for Software Artifacts (SLSA) is an end-to-end security framework designed to enhance the integrity and security of software artifacts throughout the software supply chain. SLSA offers a checklist of standards and controls for preventing tampering, securing packages, and safeguarding the infrastructure involved in software development and distribution processes~\cite{slsa, slsa-googleblog, slsa-celebrate}. In 2021, a cross-industry collaboration led to the introduction of SLSA. 

In April 2023, SLSA released v1.0, replacing v0.1 from 2021. The security requirements in SLSA are divided into tracks and levels. Each track of SLSA addresses specific aspects of supply chain security with defined requirements advancing to higher levels. While v0.1 had four tracks (source, build, provenance, common requirement) with levels 1 to 4, v1.0 focuses only on the build track with levels 0 to 3 (other tracks are deferred to future versions). The levels for both versions (0.1 and 1.0) are organized in ascending order and provide greater assurances of integrity as the level increases. 
SLSA v0.1 has four tracks; the \textit{source track} focused on code protection, requiring a version-controlled repository and trusted two-party review. The \textit{build track} emphasized securing the artifact build, including requirements like build service and hermetic builds. The \textit{provenance track} facilitated generating and consuming provenance, with requirements for availability, authentication, and non-falsifiable (now known as unforgeable). \textit{Common requirements} applied to all trusted system components, covering security, access, and superuser needs.

{SLSA} is incorporated with tools \texttt{slsa-github-
generator}~\cite{slsa-github-generator} and \texttt{slsa-verifier}~\cite{slsa-verifier}. The \texttt{slsa-github-generator} assists GitHub projects in generating and managing SLSA provenance \footnote{Provenance refers to metadata that provides information about how the outputs of a build were generated} by using GitHub actions. In addition, the {CI/CD} pipeline itself can also generate SLSA provenance as part of its build and deployment processes~\cite{GHA}. SLSA provenance is based on the in-toto~\cite{torres2019toto} attestation framework, which creates a signed document that associates metadata with a software artifact. \textit{Attestation} is an inspection process of verifying the authenticity and integrity of the generated provenance~\cite{slsa}.
The \texttt{slsa-verifier} tool performs the verification process~\cite{slsa-verifier} and verifies the SLSA provenance generated by {CI/CD} builders by checking cryptographic signatures and matching expected values such as builder ID and source code repository. 


For SLSA v1.0's build track, the requirements and objectives for each level are as follows:


\begin{enumerate}

    \item No requirements and provides no artifact integrity guarantees.  L0 represents the lack of SLSA.
    \item Requires producers to provide automatically-generated provenance that shows how a project was built (e.g., used build processes, involved entities). Producers must distribute provenance to consumers, preferably through the respective package ecosystem.
    \item Requires the build to run on a hosted platform that generates signed provenance. The downstream verification process involves validating the authenticity.
    \item Producers need to implement strong security controls on their build platform so that build runs cannot influence one another even within the same project. Also, these controls should prevent any user-defined steps from accessing secret material used to sign the provenance.
\end{enumerate}
\section{Methodology}
\begin{figure*}[h]
\centering
\includegraphics[width=.8\textwidth]{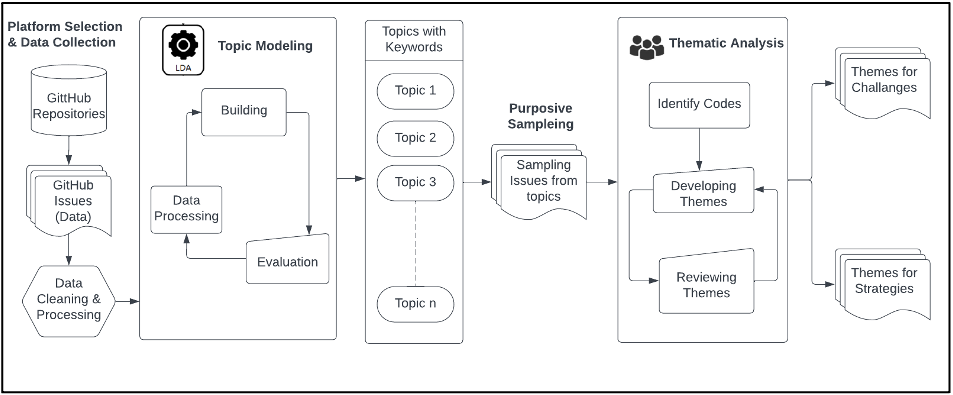}
\caption{Overview of Methodology}
\label{figure1}
\end{figure*}

Our methodology comprises five steps: Platform Selection, Data Collection, LDA topic Modeling, Purposive Sampling, and Thematic Analysis. 
 The steps in our methodology are discussed below and illustrated in Fig~\ref{figure1}.

\subsection{Data Collection Platform Selection}\label{PlatformSelection}

GitHub is a widely-used platform for software development~\cite{ITpro} and scientific research~\cite{cosentino2017systematic}. Many research articles analyze data from GitHub~\cite{imtiaz2019investigating,prana2019categorizing}. GitHub's open API and accessible features facilitate data collection and analysis, providing insights into Open-Source Software (OSS) development. SLSA hosts its project, documentation, and tools on GitHub. Authors and practitioners use GitHub issue trackers for questions, difficulties, clarifications, suggestions, and updates. Hence, we chose GitHub for data collection.

\subsection{Data Collection}
Our data collection from GitHub involved three phases: Search process, Data Cleaning and Selection, and Data Accuracy. 

\subsubsection{\textbf{Search Process}}
We applied a multi-strategy approach to compile a comprehensive SLSA-related repositories: i)Any repository within the SLSA project,
ii)Any repository that depended on \texttt{slsa-github-generator} based on the GitHub dependency graph~\cite{githubdependency}, iii)Any repository using the \texttt{slsa-github-generator} and \texttt{slsa-verifier} tools in their GitHub workflow.

\subsubsection{\textbf{Data Cleaning \& Selection}} The search phase led to collecting 733 repositories. 
We removed duplicates, and 386 repositories remained.
We reduced our initial dataset by 47\%, primarily because our third search strategy involved gathering data when the GitHub action utilized the tools.
Next, we filtered out repositories with no issues created, leaving 233 repositories collectively containing 56,992 issues. To refine our dataset, we systematically filtered these issues to isolate those directly pertinent to the SLSA framework. 
Since not all the issues gathered from the repositories referenced SLSA, we performed a keyword search using the specified search string to narrow our dataset.
Our search string keyword search involves keywords including variations, contextual keywords, specific phrases, industry jargon, and iterative feedback. The first two authors discussed and refined the strings for relevance and accuracy. Our final search strings are: ``Supply-chain Levels for Software Artifacts'', ``SLSA'', ``SLSA Framework'', ``SLSA Security Framework'', ``SLSA in Software Supply chain'', ``OpenSSF SLSA'', ``SLSA-verifier'', or ``SLSA--generator''. We included issues that contained at least one of our selected search strings in the issue's title, body, or comments. After this, 2,941 issues remained.

Next, we first discarded issues created by bots as we noticed bot-created issues were for automated tasks. 
Such bot-generated data introduces noise in communication~\cite{wessel2018power}. 
We identified if a bot created an issue using GitHub GraphQL API and checked if the issue's author type was a bot. Second, we excluded issues that contained terms indicative of automated tasks, such as branch pushing failures or scheduled workflow actions. For this, the first two authors collaboratively found some irrelevant automated issues based on any label or title included the keywords:  ``[e2e]'', ``e2e:'', ``e2e failure:'', or ``cli:'',``[cli]''. Here, ``e2e'' and ``cli'' represent end-to-end testing and command line interface. We filtered issues based on these labels since no challenges or strategies-related discussions were reported. Filtering these issues was also necessary to avoid inaccurate interference of words in LDA-generated topics. After applying these filters, 1,523 issues were left for further analysis. 

\subsubsection{\textbf{Data Accuracy}} To validate our exclusion criteria, we randomly selected 20\% of the excluded issues. The first author manually reviewed the selected excluded issues to confirm their irrelevancy and inaccuracy with the desired dataset, which bolster our method. We used the final dataset for LDA topic modeling.


\subsection{LDA Topic Modeling}
To extract underlying SLSA related topics from our dataset, we employed the Latent Dirichlet Allocation (LDA)~\cite{blei2003latent}. LDA is a widely used probabilistic algorithm for topic modeling. LDA is suitable for identifying latent patterns in textual data, such as for our study of SLSA-related issues, which encompass a wide range of concerns and complexities in data. The lack of prior, in-depth research on SLSA underscores LDA's application. LDA has been used in prior studies, such as NLP and sentiment analysis~\cite{uddin2022leveraging,uddin2023youtube,silva2021topic,gauthier2022will}.

\subsubsection{\textbf{Data Pre-processing}}
The data pre-processing stage is essential to enhance the quality of unstructured text data analysis and improve human interpretability in LDA. Our data pre-processing included: 
i) removing punctuation, numbers, stop words, white spaces, and HTML tags, and converting all text to lowercase; 
ii) applying tokenization, where our content was divided into tokens (words), which were converted into a word vector; 
iii) applying text lemmatization to reduce words to their base; and  
iv) applying n-grams (bigrams and trigrams) to capture more data context. 
We observed some common words (GitHub, issue, slsa) and acronyms repeated several times within the topics. This repetition caused the duplication of words in the topics, which led to less distinctive keywords. As such, we removed these common words from the dataset, following prior work~\cite{gauthier2022will}. We also expanded general and domain-specific acronyms, for example, 'BYOB' to Build Your Own Builder and 'sdlc' to Software Development Life Cycle. 

\subsubsection{\textbf{Building Model}}
To build the (LDA) model, we utilized MALLET~\cite{mccallum2002mallet}, which employs Gibbs sampling algorithm~\cite{geman1984stochastic}.
We iterated the model with keyword counts from 5 to 20 and topics from 5 to 60, finding the ideal model at 50 topics with 10 keywords.
We found the ideal model based on semantic consistency and the distribution of the topic-keywords based on discussion among the first two authors. 
Next, the first two authors labeled each topic separately based on the interpretation and the general meaning of the related keywords~\cite{gurcan2019big}. Then we finalized the labels and resolved conflicts by following a negotiation agreement practice~\cite{campbell2013coding} and mapped these topics into nine broader groups to facilitate the clustering of related topics~\cite{gurcan2019big}. 
The following nine labels were assigned as topic names and nine broader group names: documentation, terminology, provenance, attestation, workflow, defect management, version control, pertinence, and two-party review.

\subsection{Purposive Sampling}
For our qualitative analysis to address our research questions, we purposively sampled~\cite{campbell2020purposive} issues from each of the nine broader groups of topics to conduct our inductive coding and thematic analysis. We opted for purposive sampling to select our data, which aids in the chosen sample representing the population we are researching. We selected the top 20 issues from each group based on the composition value generated by (LDA). In the case of the ``two-party review'' group, we selected all 19 issues for our analysis because that group only contained 19 issues. As a result of our sampling process, we obtained 179 issues from 9 groups for qualitative coding. 
\begin{table*}[]\label{table1}

\caption{Categorization of themes and sub-themes of topics for challenges}
\label{table1}
\begin{center}

\begin{tabular}{|l|l|l|r|r|}
\hline
\textbf{Themes}                             & \textbf{Sub-theme}                          & \textbf{Topic groups} & \textbf{number of issues} & \textbf{number of sampled issues} \\ \hline
\multirow{4}{*}{CI. Complex Implementation} & CI.1 Complicate Provenance Generation       & Provenance            & 176                   & 20                            \\ \cline{2-5} 
                                            & \multirow{3}{*}{CI.2 Intricate Maintenance} & Workflow              & 370                   & 20                            \\ \cline{3-5} 
                                            &                                             & Defect Management     & 68                    & 20                            \\ \cline{3-5} 
                                            &                                             & Version Control       & 287                   & 20                            \\ \hline
\multirow{2}{*}{UC. Unclear Communication}  & UC.1 Unclear Definitions                    & Terminology           & 50                    & 20                            \\ \cline{2-5} 
                                            & UC.2 Unclear Documentation                  & Documentation         & 307                   & 20                            \\ \hline
\multirow{2}{*}{LF. Limited Feasibility}    & LF.1 Limited Attestation Verification       & Attestation           & 200                   & 20                            \\ \cline{2-5} 
                                            & LF.2 Two-party Review Requirements          & Two-party Review      & 19                    & 19                            \\ \hline
UR. Unclear Relevance                       & UR.1 Unclear Relevance                      & Pertinence            & 46                    & 20                            \\ \hline
\end{tabular}
\end{center}
\end{table*}


\subsection{Thematic Analysis}
 We performed a reflexive thematic analysis approach based on the six phases described by Braun and Clarke ~\cite{braun2012thematic} on purposively sampled issues. We started with inductive coding~\cite{chandra2019inductive} on selected issues for familiarization with the data and to avoid any biases toward current understanding. Inductive coding is widely utilized in academic research and involves analyzing data without a predetermined set of categories or themes~\cite{chandra2019qualitative,gauthier2022will}. The first and second authors independently reviewed all selected issues and assigned initial codes. After the initial coding, the authors met to compare and discuss generated codes and collaboratively created the final codebook  while resolving disagreements via discussion. We coded the issues from two aspects: (i) the challenges of adopting SLSA and (ii) strategies to overcome the challenges. Through an iterative process of comparing the codes, the first and second authors developed themes. Finally, to address the individual researcher positions inherent in qualitative research, such as our reflexive thematic analysis, we conducted group reflections with the first, second, and third authors on the identified themes~\cite{creswell2017designing}. 
We did not report inter-rater reliability, as this approach aligns with the principles of reflexive thematic analysis~\cite{clarke2017thematic}, and we resolved our conflicts as they emerged~\cite{wermke2022committed}. Also, the first author gathered supporting quotations for each theme.  All the study material is available in GitHub repository~\cite{githubrepo}.

\begin{figure*}[h]
\centering
\fbox{\includegraphics[width=.6\textwidth]{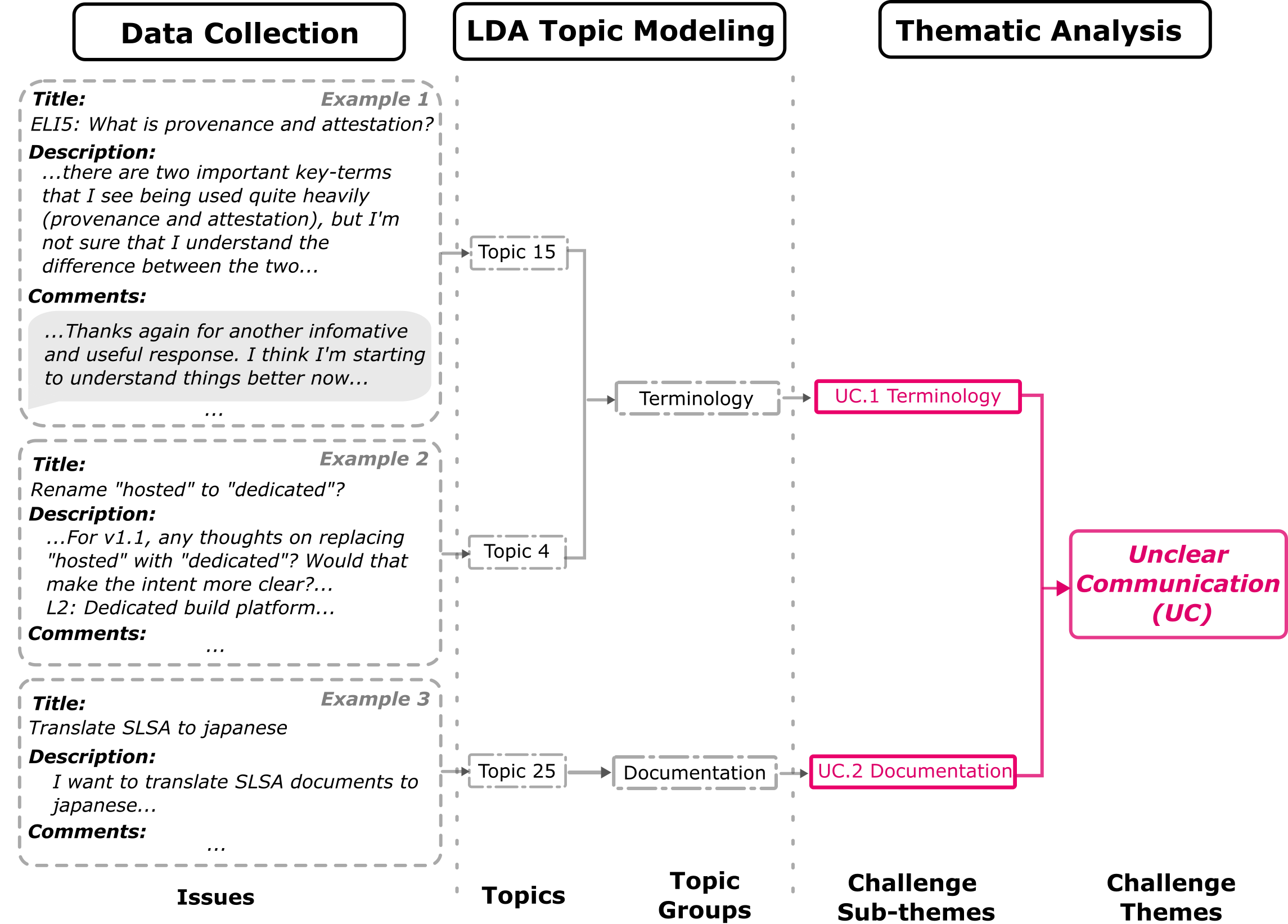}}
\caption{Processing of issues to the challenge theme}
\label{figure2}
\end{figure*}
\underline{\textbf{\textit{For challenges:}}}
The process of theming challenges is based on the type and context of the difficulties. The step-by-step process is shown in Figure~\ref{figure2}. For example, issues related to understanding SLSA-related terms were labeled as topic group `Terminology'' and grouped under sub-themes `` Unclear Definitions'' Subsequently, the sub-theme `` Unclear Definitions'' was grouped with the theme of ``Unclear Communication (UC),'' which highlights difficulties in communicating the SLSA process. Table~\ref{table1} represents the mapping of themes, sub-themes, and topic groups, along with the number of issues and sample issues.

\underline{\textbf{\textit{For strategies:}}}
Practitioners proposed strategies to address SLSA challenges, which were categorized by type, context, and method. For example, strategies for improving SLSA documentation were grouped under ``Enhance Documentation and Provide Patches." Strategies involving learning from negative experiences to improve documentation were categorized under "Use Negative Examples and Improve Design." These sub-themes were then grouped under the theme ``Provide Specific and Detailed Documentation," emphasizing the need for better SLSA documentation.

\section{Results}
In this section, we present our findings for each research question. To address RQ1, we provide challenges encountered by practitioners in Section~\ref{Subsec:challenges}. In Section~\ref{Subsec:strategies}, we provide strategies practitioners suggested to overcome these challenges to answer RQ2. Additionally, we have included de-identified quotes from the developers and practitioners gathered from the GitHub issues related to the challenge themes.

\subsection{Challenges}\label{Subsec:challenges}
For \textit{RQ1: What challenges do framework practitioners encounter
while deploying SLSA?} we identified four themes of challenges: Complex Implementation (CI)(Section.~\ref{Subsec:CI}), Unclear Communication (UC)( Section.~\ref{Subsec:UC}), Limited Feasibility (LF)(Sec. ~\ref{Subsec:LF}), and Unclear Relevancy (UR) (Section.~\ref{Subsec:UR}) from qualitative analysis. These four themes include subthemes.  We also provided the total number of issues associated with each challenge theme in parentheses.

\subsubsection{\textbf{Complex Implementation (CI)(901)}}\label{Subsec:CI} CI is related to the  challenges of integrating (SLSA) in projects. CI contains subthemes \textit{Complicate Provenance Generation} and \textit{Intricate Maintenance}.


\underline{\textbf{CI.1 Complicated Provenance Generation:}}
Practitioners reported challenges in generating provenance for higher levels of {SLSA} compliance. One major concern is the blocking nature and inefficiency of check-verifier pre-submit jobs, which can delay the submission process and affect the {CI/CD} pipeline's speed and agility. Another challenge is the lack of flexibility and support for "non-build" configurations in tools like \texttt{slsa-github-generator}. This inflexibility makes it difficult for practitioners with diverse use cases to fully utilize the tool. Additionally, some steps in using the \texttt{slsa-github-generator} can be laborious, especially with multiple builds or scripts, leading to uncertainty and difficulty in comprehension and incorporation. Security concerns also arise regarding storing sensitive information, such as usernames and login details, within the generated provenance. Ensuring secure handling of this information is important to prevent data breaches or unauthorized access. Furthermore, the risk of missing information or data validity in the generated provenance impacts the verification process.

\begin{quotation}
\textit{The integration of the slsa-framework action above is an educated guess from the instructions, but it runs. The artifact path, however, is a mystery: we've tried paths, relative paths, and the name of the generated container (tag) but are unsure exactly how to refer to the artifact}
\end{quotation}


 Standardization of the provenance generation process is essential to reduce confusion, emphasize consistency, and ensure data integrity and reliability in the build process. Semantic-release tool inconsistency in the python-package-template repository is a obstacle. The issue arises because the locally installed semantic-release tools may not match the ones used by the GitHub Action when running a CI workflow. Addressing these challenges requires technical expertise, process refinement, and adherence to security best practices throughout provenance generation and verification to meet higher levels of SLSA requirements.

\underline{\textbf{CI.2 Intricate Maintenance:}}
 Practitioners stated maintaining the specifications of the required tools of SLSA is complex. Running the \texttt{slsa-github-generator} and \texttt{slsa-github-verifier} tools causes various obstacles, including incompatibility, silent messages, runtime errors, hard coding, workflow-blocking steps, and data staleness. These technical challenges lead to confusion among users, who may be unsure about the expected behavior of the tools. Additionally, updates and modifications to the tools have sometimes resulted in discrepancies between the documentation and the actual code, complicating the maintenance process. Even minor adjustments can have unforeseen impacts, such as broken links, emphasizing the complex nature of managing project components.
\begin{quotation}
\textit{Despite a successful login to the registry, there was an issue with the cosign attest process, indicating that the problem was not related to authentication or permissions. The user had to abandon a reusable workflow and manually verify the artifact's SHA before using it}
\end{quotation}

 \begin{quotation}
\textit{Now it's really hard to see if tests are actually broken or are working fine just based on the workflow runs page. This is especially a problem because workflows fail "silently."}
\end{quotation}


Integrating multiple tracks into SLSA has led to a debate on storing each track's levels. While detailed levels provide rich information, those can complicate adoption for users. Updating project tools for improvement can introduce new bugs that require remediation or backporting. Effectively managing these changes is imperative to ensure the project aligns with each version of SLSA.
\subsubsection{\textbf{Unclear Communication (UC)(357)}}\label{Subsec:UC}
UC is related to the challenges of understanding SLSA documentation. UC contains subthemes \textit{Unclear Definitions} and \textit{Unclear Documentation}.

\underline{\textbf{UC.1 Unclear Definitions:}}
Many practitioners expressed concerns about the lack of clarity in SLSA-related terminologies, emphasizing the importance of understanding these terms for system usability. Introducing less-known terms in the SLSA framework, such as "provenance" and "attestation," has created confusion among practitioners due to the absence of defined definitions and clear meanings. This unclarity has led to ambiguity and inaccuracies in the documentation, exemplified by terms like ``hermetic build,'' ``hosted,'' ``build-service,'' and ``non-falsifiable.'' 
Inconsistent terms in SLSA documentation created confusion, especially when used interchangeably or with unclear relationships. These discrepancies with standardization between SLSA and industry underscore the necessity for clear and standardized terminologies across various ecosystems and projects.
\begin{quotation}
\textit{The words resource" and "artifact" are only briefly explained at the beginning of the "Terminology" section \& they aren't clear at all.}
\end{quotation}
\begin{quotation}
\textit{I was confused on the wording of "hosted" and asked in Slack.}
\end{quotation}
However, practitioners stated that making conventional terms is not easy in industry due to their dynamic nature and can vary based on context. Framework authors need to consider various factors while making the definitions as terminologies.

\underline{\textbf{UC.2 Unclear Documentation:}} A common obstacle the practitioners encounter is the lack of clarity in explanation and guidelines on how to apply SLSA in their ecosystem. Practitioners expressed confusion about the level requirements and the usage of variables, such as whether variables like (.Env.VERSION) require prior definition using recommended code or if they are ready to use. Additionally, inconsistencies and unclear organization in document mapping have been highlighted as a challenge. For example, discrepancies in the documentation regarding ingesting a container workflow that does not align with the current functionality.
\begin{quotation}
\textit{Need 'How to SLSA' for organizations and infrastructure providers}
\end{quotation}

\begin{quotation}
\textit{Following the documentation to migrate a project from GoReleaser, I was not able to understand that SLSA does not support "non-build" configurations}
\end{quotation}


In addition, practitioners raised concerns regarding the website design of this framework. They expressed difficulty in navigating the pages. Specifically, the attack graphs are "confusing" to interpret. Practitioners also pointed out the lack of description of the process for merging Pull-Requests. Additionally, some practitioners expressed a lack of translation of the SLSA into various languages, such as German and Japanese, to increase its global adoption.

\subsubsection{\textbf{Limited Feasibility (LF)(219)}}\label{Subsec:LF} 
 LF is related to the challenges of the feasibility of the SLSA framework for improving the software supply-chain security. LF contains subthemes \textit{Limited Attestation Verification} and \textit{Two-party Review Requirements}.

\underline{\textbf{LF.1 Limited Attestation Verification:}}
Provenance is at the core of the SLSA framework, aiding in documenting artifact authenticity. The verification process confirms artifacts' authenticity and integrity. However, practitioners are confused regarding attestation, such as its distinction from provenance and the necessity of automation in this process. Tools such as the \texttt{slsa-github-verifier} have been highlighted for complexity and redundancy, making the verification process less efficient. Moreover, a lack of clear guidance on how to communicate attestation data hinders downstream systems' ability to verify the data accurately.
\begin{quotation}{\textit{One aspect of SLSA that's still in flux is where generated attestations should be stored. Basically, there isn't really a standardized way of doing this yet? And it sounds like you're implying that Sigstore is one candidate and storing attestations within an app's own repo is an option as well, though it sounded like you're saying whether or not to do that is still an "open question".}}
\end{quotation}
\begin{quotation}
\textit{It is harder for systems to gain my trust from a producer's point of view}
\end{quotation}



 The diversity of environments for generating and storing data attestation adds flexibility but also introduces complexities and potential documentation delays. Security concerns arise regarding the accuracy of attestation due to potential bugs or vulnerabilities in the verification process. Inconsistencies between package manager registries and actual files pose another risk, potentially undermining attestation accuracy. Additionally, not all signatures meet the integrity and authenticity requirements expected by the SLSA, limiting nuanced policy decisions. Establishing trust in supply chain sources is important to overcome these challenges.

\underline{\textbf{LF.2 Two-party Review Requirements:}}
The practice of two-party review checks for every change in software component and the code approved by two qualified reviewers. This practice helps to ensure that only trusted and authenticated authorized persons can make changes in software artifacts. However, practitioners encountered challenges in implementing this practice. Identifying suitable reviewers, assessing the practice's effectiveness, and applying it across various contexts proved to be complicated. Many open-source projects have only one maintainer or active user, making finding a second reviewer challenging. In addition, implementing such a review system presents an ongoing burden in terms of time and resources, raising concerns about its cost-effectiveness. Next, practitioners discussed inconsistencies in the security requirements between "Directly submit without review" and "Modify code after review" threats. Direct submissions require a review process, whereas modifications made after an initial review do not, which leads to potential security gaps and confusion among users. 
 
\begin{quotation}
\textit{It's unclear to me if pair programming and mob programming as acceptable instances of two trusted persons.}
\end{quotation}


Concerns have emerged regarding the validity of pair programming, mob programming, and automated reviews as forms of two-person review. 
Practitioners questioned the security benefits of multiple reviews, highlighting the complexities of effective review processes.

\subsubsection{\textbf{\textit{Unclear Relevance (UR) (46)}}}\label{Subsec:UR}
 UR is related to the challenges of understanding the relevancy and significance of implementing SLSA. UR contains one sub-theme with the same name.

\underline{\textbf{UR.1 Unclear Relevance:}}
Practitioners faced confusion in identifying the specific attacks SLSA aims to mitigate and distinguishing SLSA from established security frameworks, standards, and ongoing projects. This lack of clarity makes it difficult to understand SLSA's unique benefits and value proposition. Additionally, developers struggled to determine how SLSA handles policies differently from OpenSSF best practices. For example, they were uncertain whether security policies should follow an OSSF-org policy or be managed by individual projects. This ambiguity further complicates practitioners' understanding of SLSA's distinct advantages.
\begin{quotation}
\textit{Does any SLSA level help defend against Trojan horse attacks?}
\end{quotation}

\begin{quotation}
\textit{Aids organizations in creating an inventory of software and build systems used across a variety of teams. Can we clarify this claim? How does it aid organizations? What does that mean?}
\end{quotation}


Policy variations and inconsistency in SLSA hinder its effectiveness by causing compliance complexity, resource allocation issues, and process integration challenges. For example, a discrepancy exists between the npm registry and the package manager: installing with npm install P names the package A, while using npm download P followed by npm install P.tar.gz names it B. This inconsistency affects attestation, metadata, and provenance, persisting even if resolved at build time or with a lock file. Flexibility in incorporating SLSA is also crucial for enhancing testing procedures and security policies. Practitioners struggle to understand SLSA's effectiveness, particularly in creating a software repository for multiple teams, which hampers software management and collaboration.
\subsection{Strategies}\label{Subsec:strategies}
For \textit{RQ2:What strategies do software practitioners suggest to framework authors for increasing SLSA adoption?}, we have analyzed strategies suggested by practitioners to overcome the challenges and identified five main themes and 13 sub-themes.

\subsubsection{S1. Enhance SLSA alignment and flexibility}
\hspace{2pt}

\underline{S1.1 Incorporate Build-System Tracks:} 
 A "build systems track" should be included, focusing on the security of build systems in addition to the existing "build track" and "source track." Incorporating build system tracks within the SLSA framework allows for more tailored approaches to diverse system requirements. While current material on verifying build systems is a good start, it has limitations, such as the inability of people to self-assess a build system and the lack of visibility into how a provider performs.
 
\underline{S1.2 Gamify the Environment:} 
Consumers want to know at a glance. Producers aim for higher levels as a form of gamification that can help them make their systems more secure. To provide rich information about the security of consumers, gamification can aid in promoting better adoption and comprehension of SLSA.

\underline{S1.3 Ensure Flexibility:} Maintaining flexibility within the SLSA framework is essential to accommodate diverse system demands. The flexibility involves providing customizable and adaptable options based on specific organizational needs and evolving security landscapes. Furthermore, internal use cases can be different from what the open-source community needs. 

\underline{S1.4 Align SLSA with OpenSSF Best Practices:} 
Practitioners suggested adding SECURITY.md, which at least points to the OpenSSF policy. The security policy should also identify the "security team" of members who are knowledgeable about security and will address security issues in order to better comply with OpenSSF security best practices. Adopting OpenSSF best practices silver and gold criteria, adding detailed reference docs for each action and workflow, and implementing an organization-wide security policy for the \texttt{slsa-github-generator} will enhance SLSA's effectiveness. 
\subsubsection{S2. Provide specific and detailed documentation}
\hspace{2pt}
\\
\underline{S2.1 Enhance Documentation and Provide Patches:}
Improving the documentation involves providing clear and robust definitions, revising terms, using standard terms, and rewriting the get started page. The requirements need to be more precise and aligned with the SLSA levels to make it easy to apply them to any build system without any significant modifications.

\underline{S2.2 Use Negative Examples and Improve Diagram:}
Practitioners suggested the use of negative examples to explain complex concepts, define the framework's scope, or describe the purpose of requirements. This approach ensures an efficient and effective software management workflow while improving user experience and system functionality.
Furthermore, practitioners have emphasized improving website design and incorporating diagrams to enhance user experience. For example, linking terminology pages with supply chain diagrams can provide a more precise understanding.

\subsubsection{S3.Streamline Provenance Generation Processes}
\hspace{2pt}

\underline{S3.1 Simplify and Standardize Provenance Generation:}
Standardizing the provenance generation process to enhance data integrity and reduce confusion and emphasizing consistency and reliability in builds are suggested. This approach simplifies and standardizes processes with tools and templates, making it easier for application owners to adopt provenance in development and deployment.

\underline{S3.2 Fix semantic-release tool inconsistency:} To deal with semantic tool discrepancy in provenance generation, practitioners suggested defining clear rules for versioning based on semantic guidelines and using tools. Additionally, providing proper documentation and training to align teams and regularly test the release tool to catch issues early. Optimizing pre-submit jobs, such as running tasks in parallel, can also improve efficiency and consistency in releases.

\subsubsection{S4.Improve SLSA Verification Process}
\hspace{2pt}
\\
\underline{S4.1 Strengthen Verification Processes:} To strengthen the verification process practitioners proposed enhancing security guarantees, providing algorithms for determining artifact levels, and offering additional evidence of verification. Implementing these strategies improves reliability, accuracy, and security in the verification.

\underline{S4.2 Implement Versioning Tagging:}
Practitioners emphasized the importance of
implementing versioning tagging during the early stages of the SLSA framework to facilitate more straightforward tracking of
progress and changes.

\underline{S4.3 Enhance SLSA Framework and Tool:}
Practitioners proposed enhancements to the SLSA framework and tool by adding more signaling information for downstream users. The strategy will improve the overall functionality and usability of SLSA.

\subsubsection{S5.Collaborate with Community}
\hspace{2pt}
\\
\underline{S5.1 Foster Community Engagement:} 
Collaborating within communities aids in enhancing security measures~\cite{amft2024everyone}. Aligning SLSA verification practices with industry standards and guidelines promotes industry-wide compatibility and adoption. Providing clarification and explanations will aid novice users in understanding new technologies. Emphasizing community engagement will help in enhancing framework usability, leading to improved user experiences.

\underline{S5.2 Promote Learning and Knowledge-Sharing:} Practitioners and framework authors have fostered a culture of learning and knowledge-sharing. Their collaboration is evident in addressing challenges like the two-party review, proposing solutions for single maintainers, low funding, and fewer requirements. For example, single maintainers can collaborate with other single-author projects for mutual support and expertise exchange.

\section{Discussion}

 We discuss the mapping between challenges and strategies and make recommendations for security framework authors (Section.~\ref{SUBSEC:authors}), practitioners (Section.~\ref{SUBSEC:practicioners}), and researchers (Section.~\ref{SUBSEC:researchers}).

\subsection{For security framework authors}~\label{SUBSEC:authors}
 Our findings highlight that Unclear Communication (UC) is a primary concern, leading to disinterest or confusion among adopters. To address the Unclear Definition challenge  (UC.1), we recommend standardizing and consistently defining terminology, providing clear examples, and developing a comprehensive, easily accessible glossary. To deal with Unclear Documentation (UC.2), document quality can be improved by implementing strategies such as enhancing documentation and providing patches (S2.1), incorporating negative examples, improving diagrams (S2.2), increasing traceability, and considering translations. Framework authors should highlight the framework's uniqueness by contrasting it with other approaches and enhancing SLSA's alignment and flexibility (S1) to address the issue of Unclear Relevance (UR). Clear relevancy can be achieved by incorporating build-system tracks (S1.1), gamification of the environment (S1.2), ensuring flexibility (S1.3) and aligning SLSA with OpenSSF best practices (S1.4). These efforts collectively contribute to showcasing the framework's utility and increasing its appeal to potential adopters. Furthermore,  
Complex Implementation (CI) was found to be a significant challenge. To simplify this, the data suggests extending tools to improve the SLSA framework and tools (S4.2), fixing semantic-release tool inconsistency (S3.2) to deal with the discrepancy, simplifying and standardizing provenance generation (S3.1), and enhancing the SLSA framework and tool (S4.3).

We recommend framework authors enhance the SLSA framework by i) improving documentation by providing detailed guidelines, templates, and comprehensive examples; ii) implementing user-friendly strategies, such as designing intuitive interfaces and interactive demos; iii) automating processes with flexible tools by developing configurable tools for provenance generation and artifact verification, and integrate continuous security monitoring.

\subsection{For practitioners}~\label{SUBSEC:practicioners}
When adopting SLSA, practitioners should be aware of several challenges. First of all, understanding and ensuring that the framework implementation aligns with the project's security needs and goals
Due to concerns about Limited Feasibility (LF), practitioners should carefully study and verify SLSA's security checks. For instance, verifying an attestation (LF.1) verifies specific build steps with particular inputs that lead to certain outputs. Weak links in certain package managers, such as those from npm, can potentially compromise the effectiveness of the attestation process ~\cite{zahan2022weak}. Moreover, the two-party review (LF.2) might exclude certain projects when integrated into future framework versions. Practitioners should be aware of these limitations and plan their security strategies.

Despite the challenges, we recommend practitioners integrate SLSA and actively contribute to the improvement of the security framework to foster community engagement (S5.1) and promote learning and knowledge-sharing (S5.2). By engaging in collaborative efforts and sharing insights for enhancements, practitioners can play a vital role in evolving SLSA to address emerging challenges and strengthen its effectiveness across diverse software development environments. As such, always giving back to projects is vital for the sustainability of the ecosystem~\cite{wermke2023always}. Practitioners can contribute by joining Slack to discuss with fellow developers, participating in community meetings, and improving SLSA with GitHub issues~\cite{slsacommunity}.
 We recommend practitioners adopt the SLSA framework while carefully verifying security checks, understanding limitations, and ensuring its relevance to the project's security needs. Actively contribute to the SLSA community by sharing insights, proposing enhancements, and participating in discussions to help evolve the framework and enhance its effectiveness.
\subsection{For researchers}~\label{SUBSEC:researchers}
Several software supply chain challenges exist that require further research and development efforts. Practitioners often find it difficult to trust code that was not developed by themselves~\cite{williams2022trusting}, which is essential in the correct generation of provenance (CI.1). Ensuring all provenance-building activities occur within the threat model's trust boundary is important but complicated for users. Additionally, trust in the attestation (LF.1) is necessary.
Inaccuracies, limitations, or vulnerabilities can harm the trust in the verification process and the actor. Practitioners mentioned possible exploits, including tampering with environments for pair programming, malicious collaborators, changing reviewed code, and subverting tools. To overcome the challenges, trust can be achieved by strengthening verification processes (S4.1) and implementing versioning tagging (S4.2). More research into mechanisms, tools, and standards to build trust throughout the supply chain is needed to increase the reliability of software packages~\cite{slsa-lunch}. Another challenge is the sustainability of open-source software projects when adopting SLSA practices. The two-party review process (LF.2) aims to balance security with the practicality of implementation for open-source, but many projects have few or single maintainers, making them susceptible to attacks. Two-party review can provide better guarantees but may limit accessibility and easy adaptability, which is a key motivator for practitioners. Research is required on lowering adoption barriers, understanding contributor motivation, and mitigating disengagement. Studies can focus on detecting systems that are at risk through measurement~\cite{calefato2022will, zahan2023openssf}, easing adoption barriers for newcomers~\cite{balali2018newcomers}, addressing natural disengagement within projects~\cite{miller2023we} and understanding and improving developer motivation to contribute~\cite{gerosa2021shifting}. In addition, automating SLSA processes requires further work. While tools like \texttt{slsa-github-generator} and \texttt{slsa-github-verificator} aid adoption for provenance generation (CI.1) and verification (LF.1), more tooling is still needed.

\section{Threat To Validity}
In our study,  we did not collect issues from other platforms, such as Reddit ~\cite{reddit} and Stackoverflow~\cite{stackoverflow}. We recognize the limitation by considering the widely adopted and recognized platform, Github, in software engineering and security studies ~\cite{almarzouq2020mining, pew,usinggithub}. Next, we did not collect demographic information such as gender, age, occupation, and technical background of practitioners. As such, we cannot assess the experience and perspectives of practitioners in the SLSA community. Moreover, this limitation restricted our capacity to generate descriptive statistics, such as tracking the number of issues created by individual users or investigating interactions between community members.  
We accounted for limitations in our topic modeling approach while optimizing models. Following Baumer et al.\cite{baumer2017comparing}, we prioritized human-interpretable models, focusing on providing insightful data perspectives over seeking the optimal model. Our methodology, guided by LDA analysis practices\cite{maier2021applying}, included thorough data cleaning and selecting topic numbers based on semantic coherence and keyword distribution. We also applied n-grams and word embedding to overcome the bag of word assumption of LDA. Finally, varying interpretations and potential oversights, manual analysis may introduce bias. For instance, the identified categories of challenges and strategies are susceptible to such bias. To address this, we cross-checked the identified categories and included only those on which both authors agreed.
\section{Ethical Consideration}
Our Institutional Review Board (IRB) classified this study as ``not human subjects'' as we only utilized data that was publicly accessible and did not include interaction with humans. We also followed GitHub's terms of use and guidelines for running academic research. According to GithHub's Acceptable Use Policies~\cite{github:2023}. We did not collect usernames or email addresses, so we will not publish any user information. We will publish only aggregate information and short, pseudonymized quotes from GitHub posts.
\section{Related Work}
The software supply chain has become a frequently targeted attack vector in the field of cyberattacks~\cite{liu2022demystifying}. For both open and closed-source supply chains, ensuring the reliable and efficient operation of their systems is essential ~\cite{raymond1999cathedral} as it involves compromising downstream dependents ~\cite{ohm2020backstabber} and external factors~\cite{du2013towards} in different ecosystems~\cite{zimmermann2019small}. Lella et al.~\cite{lella2021enisa} presented a taxonomy of supply chain attacks outlining the techniques used by attackers and the targeted assets. Ladisa et al.~\cite{ladisa2023sok} classified attacks on open-source supply chains across all stages. Williams et al.~\cite{williams2024proactive} proposed the Proactive Software Supply Chain Risk Management (P-SSCRM), a comprehensive framework for organizations to proactively manage software supply chain risks. 
Proposed solutions like The Update Framework (TUF)~\cite{samuel2010survivable}, in-toto~\cite{torres2019toto} help to ensure supply chain integrity and secure distribution.
In addition, tools and methods such as CDI~\cite{CDI-contibutors}, SPIRE~\cite{SPIRE-contibutors}, and Sigstore \cite{Sigsore} provide software signing capabilities for developers, minimizing adoption barriers. Security tools protect package users by identifying and addressing known dependencies, but infrastructure is needed to build the framework.\cite{liu2022demystifying,wyss2022fork,okafor2022sok,melara2022software}. Merala and Bowmen~\cite{melara2022software} stated, based on use case studies, that SLSA helps establish trust, enables trust flow between entities, and helps to ensure supply chain security. SLSA compliance with software bill of materials (SBOM)~\cite{SPDX-contibutors} to gain package information, SLSA can enhance software resiliency against potential attacks on the supply chain ~\cite{Cycode}~\cite{SPDX-contributors}. According to Enck and Williams~\cite{enck2022top} findings from three industry summits, experts have a positive attitude towards SLSA, but securing the built environment remains challenging due to issues like trusting the compiler, among others.  Organizations are adopting SLSA to fortify infrastructure, yet hurdles remain~\cite{tran2023s3c2,dunlap2023s3c2}; drawbacks across software security frameworks lead
to a lack of implementations ~\cite{Garantir,kalu2024industry,zottmann2023comparing} underscoring the necessity of addressing adoption challenges and enhancing the framework's efficacy. Our research is inspired by previous studies, focusing on uncovering challenges faced by practitioners when deploying SLSA. We also explore strategies proposed by practitioners to help developers address challenges and improve SLSA adoption.
\section{Conclusion}
Software security frameworks, such as SLSA, are designed to aid in securing projects throughout the software supply chain. However, despite the growing interest in adopting SLSA, practitioners face challenges. To understand the challenges and to find effective strategies to mitigate them, we conducted a content analysis of SLSA-related issues on GitHub. We analyzed 1,523 issues from 233 software repositories and leveraged probabilistic topic modeling (LDA) to identify latent topic sampling issues for qualitative analysis. Through thematic analysis, we identified four themes representing the challenges and five strategies to address them.

Our analysis revealed the top challenges, with the highest number of reported issues related to complex implementation and unclear communication of the SLSA process. The suggested strategies to address these challenges include streamlining provenance generation processes, improving the SLSA verification process, and providing specific and detailed documentation to overcome the challenges. Our findings emphasize the recurring need to simplify the implementation and understanding of security frameworks while enhancing trust in software supply chain security. Effective collaboration among framework authors, researchers, and practitioners is essential to improving adoption rates and strengthening software supply chain security. 
\section*{Acknowledgment}
This work was supported and funded by National Science Foundation Grant No. 2207008. Any opinions expressed in this material are those of the author(s) and do not necessarily reflect the views of the National Science Foundation.

\bibliographystyle{IEEEtran} 
\bibliography{main}

\end{document}